\documentclass[plain, 12pt]{article}
\usepackage[vmargin=1.8cm, hmargin=3.2cm]{geometry}
\UseRawInputEncoding

\RequirePackage[OT1]{fontenc}
\usepackage{amsthm,amsmath,amssymb,enumitem,authblk,bm,graphicx}

\usepackage[english]{babel}  

\usepackage{tikz}
\usetikzlibrary{arrows.meta}

\usepackage{setspace}
\usepackage[round]{natbib}

\usepackage{tikz}
\tikzset{>={Latex[width=2mm,length=2mm]}}

\newcommand{\E}{\mathrm{E}}
\newcommand{\R}{\mathbb{R}}

\newcommand{\re}{\mathbb{R}}
\newcommand{\eg}{{\it e.g.}}
\newcommand{\ie}{{\it i.e.}}
\newcommand{\deffeq}{\mathrel{\overset{\makebox[0pt]{\mbox{\normalfont\tiny\sffamily def}}}{=}}}

\newcommand{\tth}{^{\mbox{th}}}
\newcommand{\logit}{\mbox{logit}}
\newcommand{\logitInv}{\mbox{logit}^{-1}}
\newcommand{\dd}{^{[d]}}

\newcommand{\Po}{\mbox{Po}}
\newcommand{\Bi}{\mbox{Bi}}

\newcommand{\Fnine} {\fontsize{9}{11}\selectfont  }
\newcommand{\Ften} {\fontsize{10}{11}\selectfont  }

\renewenvironment{abstract}
{\small
  \begin{center}
  \bfseries \abstractname\vspace{-.5em}\vspace{0pt}
  \end{center}
  \list{}{
    \setlength{\leftmargin}{.0cm}%
    \setlength{\rightmargin}{\leftmargin}%
  }%
  \item\relax}
{\endlist}

\usepackage[section]{placeins}

\usepackage{booktabs, siunitx}

\begin{document}

\pagestyle{myheadings}
% \markboth{Multivariate Generalised Linear Mixed Models for Studying Roots Development}
\markboth{J.S. Pelck and R. Labouriau}
{J.S. Pelck and R. Labouriau}
\thispagestyle{empty}

\title{\Large
Using Multivariate Generalised Linear Mixed Models \\
for Studying Roots Development: \\ \large
An Example Based on Minirhizotron Observations.
}

\author[1] {Jeanett S. Pelck}
\author[1,] {Rodrigo Labouriau 
  \thanks{Corresponding author: Rodrigo Labouriau, rodrigo.labouriau@math.au.dk}}
\affil[1]{Department of Mathematics, Aarhus University, Denmark}
\date{November 2020}

\clearpage\maketitle
\thispagestyle{empty}

\begin{abstract}\Ften
The characterisation of the spatial and temporal distribution of the root system in a cultivated field depends on the soil volume occupied by the root systems (the scatter), and the local intensity of the root colonisation in the field (the intensity). We introduce a multivariate generalised linear mixed model for simultaneously describing the scatter and the intensity using data obtained with minirhizotrons (i.e., tubes with observation windows, which are inserted in the soil, enabling to observe the roots directly). The models presented allow studying intricate spatial and temporal dependence patterns using a graphical model to represent the dependence structure of latent random components.

The scatter is described by a binomial mixed model (presence of roots in observation windows). The number of roots crossing the reference lines in the observational windows of the minirhizotron is used to estimate the intensity through a specially defined Poisson mixed model. We explore the fact that it is possible to construct multivariate extensions of generalised linear mixed models that allow to simultaneously represent patterns of dependency of the scatter and the intensity along with time and space.  

We present an example where the intensity and scatter are simultaneously determined at three different time points. A positive association between the intensity and scatter at each time point was found, suggesting that the plants are not compensating a reduced occupation of the soil by increasing the number of roots per volume of soil. Using the general properties of graphical models, we identify a first-order Markovian dependence pattern between successively observed scatters and intensities. This lack of memory indicates that no long-lasting temporal causal effects are affecting the roots' development. The two dependence patterns described above cannot be detected with univariate models.\end{abstract}


\normalsize
% \newpage

\newgeometry{top=3.5cm, bottom = 2cm, right = 3cm, left = 3cm}

\newpage 

\tableofcontents

\newpage

\section{Introduction}
The characterisation of the spatial and temporal distribution of the root system of a cultivated field depends, among other factors, on two key features: 
the volume occupied by the root systems in the field, here called the \emph{scatter} (or root \emph{frequency} in the terminology of \citealp{Kristensen2004}), and the local intensity of the root colonisation in the field, termed \emph{intensity}. These two characteristics might vary with time and the use of different cultivation practices or treatments (see \citealp{Kristensen2004,Kristensen2007,Kristensen2017,Hefner2019,Christensen2020etal}). This article introduces and discusses a multivariate statistical model for simultaneously describing the root \emph{scatter} and  \emph{intensity} using data obtained with a device called minirhizotron (briefly described below). 
Additionally, the models presented will allow us to study intricate spatial and temporal dependence patterns using a version of the so called \emph{graphical model}, representing the dependence structure of latent random components.

A minirhizotron consists of a tube along with there are several transparent observation windows. According to the methodology, several tubes are inserted in the soil, allowing for observing the development of the roots at several depths, positions, and time points in the field. After a given growth period, the observation windows are examined using a camera introduced in the tube, and the presence or absence of roots in each window is registered. Each observation window has reference lines and the number of times the roots cross (if any) the reference lines are also recorded. 

The main idea explored here is that the presence or not of roots in the observation windows is the result of sampling in the field and, therefore, can be used to characterise the volume occupied by the root system in the field, \ie, to quantify the \emph{scatter}. Furthermore, we will argue that the number of times the roots cross the reference lines can be used to estimate the \emph{intensities}.
% length of the roots that occupy (a slice of) a region of the soil around each observation window, which is a quantity directly related to the local root intensity.

The \emph{intensity} and \emph{scatter} will both be modelled using suitable generalised linear mixed models, as described below. The \emph{scatter} will be described by a binomial model (presence or not of roots in observation windows). The number of roots crossing the reference lines will be modelled using a specially defined Poisson model. A stochastic geometric argument will allow us to use the number of crosses to obtain estimates of the length of the root system in the region surrounding the observation windows of the minirhizotrons. The models used here will contain random components, 
% (therefore are classified as "mixed" models)
allowing to represent the dependence structure induced by the experimental designs typically used in the applications we have in mind. 

We will explore the fact that it is possible to construct multivariate extensions of generalised linear mixed models that allow to simultaneously represent, in a single model, patterns of dependence of the \emph{scatter} and the \emph{intensity} along with time and space. This achievement is remarkable since the nature of these two quantities is very different. We will show an example where the \emph{intensity} and the
\emph{scatter} are simultaneously determined at three different time points. Jointly modelling these six quantities will allow us to identify a first-order Markovian dependence pattern between successively observed \emph{scatters} and \emph{intensities}. Moreover, we will show a positive association between the \emph{intensity} and the \emph{scatter}, quantify the magnitude of those associations at the three different observation times, and show that there is a decay of the association between  the \emph{intensity} and the \emph{scatter} at the last observation time. This type of characterisation of the time development of the root system cannot be obtained using only univariate analyses.

This article is organised as follows. 
Section \ref{SubSection.2.1} presents a motivational real example. 
The basic models for the \emph{scatter} and the \emph{intensity} are discussed in the sections  \ref{SubSection.2.2} and \ref{SubSection.2.3}, respectively. In section \ref{Section.3} we introduce a multivariate model for describing the roots' colonisation at different developing stages simultaneously. After defining a multivariate generalised linear model connecting the univariate models 
% through random components defined for taking the same values in the same experimental units 
in section \ref{SubSection.3.1}, we model the covariance structure of those random components using a graphical model in section  \ref{SubSection.3.2}, and briefly describe techniques for inferring this graphical model in section \ref{SubSection.3.3}. The motivational example presented in section \ref{SubSection.2.1} is analysed in section \ref{Section.4} using the multivariate model described. Section \ref{Section.5} presents a brief discussion of the methods exposed.

%%%%%%%%%%%%%%%%%%%%%%%%%%%%%%%%%%%%%
%%%%%%%%%%%%%%%%%%%%%%%%%%%%%%%%%%%%%
\section{Models for Scatter and  Intensity of the Roots' Colonisation}\label{Section.2}

%%%%%%%%%%%%%%%%%%%%%%%%%%%%%%%%%%%%%
\subsection{A Motivational Reference Example}\label{SubSection.2.1}

We consider below a real example arising from a study on the effects of different liming and phosphorous fertilisation techniques in a field experiment (see \citealp{Christensen2017} and  \citealp{Christensen2020etal}). 
This example will be used to expose the modelling approach studied in this article.
In this study, an experimental field cultivated with spring barley was split into three blocks containing four plots; in each block, four fertilisation treatments were randomly allocated to the plots. In each plot, two minirhizotron tubes were installed. Three soil depth zones were considered in the analyses below: the superficial layer (termed horizon A), the intermediate layer (called horizon B), and the subsoil (termed horizon C). The minirhizotron tubes had six observation windows in the superficial layer and twelve windows in the other two layers. The observation windows of all the 24 tubes were examined in three time points corresponding to different development stages of the culture of spring barley (see details in \citealp{Christensen2017}). For simplicity of the exposition, we ignore the block and plot structure of the experiment. 

The primary interest in the study referred above was to characterise the development of the root system in each soil depth zone when different fertilisation treatments are used. Here we approach a different question of characterising how the dependence between the \emph{intensity} and \emph{scatter} vary over the three observed development stages of the culture in the field. This problem involves studying the dependence of quantities of different stochastic nature. Indeed, while we will characterise the \emph{intensity} using the counts of number of times the roots cross the reference lines in the observation windows, the \emph{scatter} will be characterised examining the incidence of roots in observational windows. 

The strategy we will adopt to analyse this example is to construct suitable multivariate generalised linear mixed models describing the rooting \emph{intensities} and the \emph{scatters} at the three developmental stages (so the model will be six-dimensional). The one-dimensional generalised linear mixed models describing these two characteristics of the rooting system were first described in \cite{Labouriau2019} and are presented in detail in section \ref{Section.2}.  The idea we will explore is that the fixed effects of the models will adjust for the expected differences due to the treatments and the soil depth zones. Each of these models will contain a Gaussian random component taking the same value for each observation arising from the same minirhizotron tube (here we interpret the tubes as the experimental units). Those random components represent the local variation of the \emph{intensity} or the \emph{scatter} present at each experimental unit after having corrected for the effects of the depth zones and the fertilisation treatments. The multivariate model we will consider will allow us to represent different covariance structure of the six Gaussian random components corresponding to the six observed responses. The covariance structure of the random components will determine the covariance structure of the responses, as we discuss below.
 
 %%%%%%%%%%%%%%%%%%%%%%%%%%%%%%%%%%%%%
\subsection{Modelling the Scatter}\label{SubSection.2.2}

 We model the \emph{scatter} at a fixed development stage by studying the occurrence of roots in the different observational windows of the minirhizotron tubes. Denote by $Y_{tkz}\dd$ the random variable representing the number of windows where a root is present in the $z\tth$ soil depth zone ($z=A, B, C$ representing the soil horizons) at the $k\tth$ tube ($k=1,\ldots,6$) exposed to the $t\tth$ treatment  ($t=1,\ldots, 4$), observed at the $d\tth$ development stage ($d=1,2,3$). We keep the development stage fixed in this section and in Section \ref{SubSection.2.3}. Moreover, following the same convention for the sub-indices used above, denote the number of observational windows at the $tkz\tth$ observation by  $n_{tkz}$. Note that by design, the number of observation windows does not change in the different observation times.

Denote, for $t=1,\ldots , 4$ and $k=1,\ldots,6$, by $U_{tk}\dd$ an unobservable random variable taking the same value for all observations arising from the $tk\tth$ tube. We assume that those random variables, corresponding to the $24$ tubes used in the experiment, are independent and normally distributed with expectation zero and variance $\sigma^2_{U[d]}$. According to the model in discussion, the random variables $Y_{1A1}\dd, \dots , Y_{4C3}\dd$, representing the observations, are conditionally independent given the random components $U_{11}\dd, \dots, U_{46}\dd$. Moreover, we assume that, for $t=1,\ldots, 4$, $k=1,\ldots,6$ and $z=A, B, C$, the random variable $Y_{tkz}^d$ is conditionally binomially distributed  given $U_{tk}\dd$, with
$Y_{tkz}\dd\vert U_{tk}\dd = u\sim \Bi( n_{tkz}, p_{tkz}\dd)$, where
\begin{align}\label{Eq3.2.1}
	\logit(p_{tkz}\dd)= \beta_{tz,[d]}+u, \mbox{ for all } u \in \re \, .
\end{align}
The model described above coincides with a generalised linear mixed model (GLMM) defined with the
binomial distribution, the logistic link function, a fixed effect representing the interaction of
treatment and soil depth zone, and a random component representing the tubes. 

The parameter $\beta_{tz,[d]}$ in (\ref{Eq3.2.1}) is clearly related to the \emph{scatter}. 
Indeed, according to the model above, the probability of finding a root which is visible in an observation window at the $z\tth$ soil depth of the plots that received the $t\tth$ treatment ($z = A,B,C$ and $t=1,\dots,4$)  at the $d\tth$ development stage is
\begin{align}\label{Eq3.2.2}
  E \Big [\frac{Y_{tkz}\dd}{n_{tkz}} \Big ] =
  \int_\R \frac{\exp (\beta_{tz,[d]} + u )}{1+ \exp (\beta_{tz,[d]} + u)} \varphi (u; 0, \sigma^2_{U[d]}) du    \deffeq
  \alpha_{tz}(\beta_{tz,[d]},  \sigma^2_{U[d]}) \deffeq
  \tilde\alpha_{tz}\dd \, .
\end{align}
Here $\varphi(\cdot;0,\sigma^2_{U[d]})$ denotes the density of  a normal distribution with  expectation $0$ and variance $\sigma^2_{U[d]}$, which is the distribution of the random component $U_{tk}\dd$. 
The quantity $\tilde\alpha_{tz}\dd$ can easily be evaluated once we have estimated the parameters $\beta_{tz,[d]}$ and $ \sigma^2_{U[d]}$ by numerically integrating the integral in  (\ref{Eq3.2.2}) or using a straightforward Monte Carlo integration.  

 %%%%%%%%%%%%%%%%%%%%%%%%%%%%%%%%%%%%%
\subsection{Modelling the Intensity}\label{SubSection.2.3}

Let $C_{tkz}\dd$ be a random variable representing the total number of times the roots cross the reference lines in all the observational windows at the $z\tth$ soil depth zones ($z=A, B, C$) at the $k\tth$ tube ($k=1,\ldots,6$) exposed to the $t\tth$ treatment  ($t=1,\ldots, 4$), observed at the $d\tth$ development stage ($d=1,2,3$, fixed along this section).

Denote, for $t=1,\ldots,4$ and $k=1,\ldots,6$, by $V_{tk}\dd$ an unobservable random variable taking the same value for all observations arising from the same tube. Those random variables are assumed to be independent and normally distributed with expectation zero and variance $\sigma^2_{V[d]}$. The random components defined above are analogous to the random components used for modelling the \emph{scatter}. According to the model, the random variables $C_{1A1}\dd, \dots , C_{4C3}\dd$, representing the observations of numbers of crosses, are conditionally independent given the random components $V_{1 \, 1}\dd, \dots, V_{4 \, 6}\dd$. Moreover, we assume that, for $t=1,\ldots, 4$, $k=1,\ldots,6$ and $z=A, B, C$, the random variable $C_{tkz}\dd$ is conditionally Poisson distributed  given $V_{tk}\dd$, with conditional expectation given by
\begin{align}\label{Eq3.3.1}
	\log(\E[C_{tkz}\dd \vert V_{tk}\dd= v])= \theta_{tz,[d]}+ v +\log \left ( n_{tkz} \right ) \mbox{ for all } v\in \re \, .
\end{align}

The model above allows us to estimate the local length of the roots visible in the windows, characterising in this way the root \emph{intensity}, as described below.  
Exponentiating both sides of (\ref{Eq3.3.1}) and taking expectations with respect to the distribution of the random components yields for $t=1,\ldots,4$, $k=1,\ldots,6$ and $z=A,B,C$, that
\begin{align}\label{Eq3.3.2}
	 E \Big [\frac{C_{tkz}\dd}{n_{tkz}} \Big ] =
	 \int_{\R} \exp(\theta_{tz,[d]})\exp(v) \varphi (v; 0, \sigma^2_{V[d]}) dv =
	 \exp(\theta_{tz,[d]}) \exp(\sigma^2_{V[d]}/2)  \deffeq
	  \omega_{tkz}\dd \, .
\end{align}
The factor $\exp(\sigma^2_{V[d]}/2) $ in the right side of (\ref{Eq3.3.2}) is the expectation of the corresponding log-normal distribution (see \citealp{AitchisonBrown1957}). 

The quantity $\omega_{tkz}\dd$ defined in (\ref{Eq3.3.2}) is straightforwardly related to the \emph{intensity} since the more intense the root colonisation process in a region around the tube is, the more likely will be the occurrence of roots crossing the reference lines in the observational windows. 
Additionally, $\omega_{tkz}\dd$ can be interpreted as an estimate of the length of the roots that are visible in an observational window using the argument sketched below. A classic argument for the Buffon's needle problem allows one to calculate the length of a rigid straight needle by randomly throwing the needle in a surface with parallel reference lines (see \citealp{KlainRota1997}), the length of the needle being proportional to the probability of the needle cross a line.  
The Buffon's needle problem can be extended by dropping the assumption that the needle has a perfect straight form, yielding the so called Buffon's noodle problem. According to \cite{Ramaley1969} the length of a one dimensional structure (the "noodle" replacing the needle) is proportional to the mean of the number of times the structure crosses the reference lines. Taking this approach, the left size of (\ref{Eq3.3.2}) is interpreted as the expected value of the Buffo's noodle estimate of the length of the roots that are visible in the observation windows.

The model described above coincides with a generalised linear mixed model (GLMM) defined with the
Poisson distribution, the logarithm link function, a fixed effect representing the interaction of
treatment and soil depth zone, an offset representing the logarithm of the number of observational windows and a random component representing the tubes.

%%%%%%%%%%%%%%%%%%%%%%%%%%%%%%%%%%%%%
%%%%%%%%%%%%%%%%%%%%%%%%%%%%%%%%%%%%%
\section{Multivariate Simultaneous Models for  the Scatter and  the Intensity of the Roots' Colonisation}\label{Section.3}

%%%%%%%%%%%%%%%%%%%%%%%%%%%%%%%%%%%%%

\subsection{The Multivariate Construction}\label{SubSection.3.1}

We introduced GLMMs representing, separately, the \emph{scatter} and the \emph{intensity} of the root colonisation at a given development stage in Section \ref{Section.2}. Here, we construct a multivariate model for simultaneously describing these two characteristics at the three observed development stages. First, we define a GLMM for the \emph{scatter} and  for the \emph{intensity} for each of the three development stages, as we explained above. In each of these models, there is a random component taking the same value for all the observations arising from the same experimental unit (\ie, the same tube). Therefore,  we might connect the models by assuming that the six random components are multivariate normally distributed. As we will argue below, the covariance structure of the multivariate distribution of the random components will allow us to characterise a type of association between the \emph{scatters} and the  \emph{intensities} observed in the same or at different development stages. The details of this construction are given below.

According to the multivariate GLMM that we propose, for  $t=1,\ldots, 4$, $k=1,\ldots,6$, $z=A, B ,C$, and $d=1,2,3$,
\begin{align}\label{Eqn.3.1.1}
\left \{
 \begin{array}{l}
    Y_{tkz}^{[1]} \vert U_{tk}^{[1]} = u_1 
     \sim \Bi \left (n_{tkz},  \logitInv \left \{ \beta_{tz, [1]} + u_1  \right \}  \right ) 
      ,\, \forall u_1\in\re 
     \\
          Y_{tkz}^2 \vert U_{tk}^2 = u_2 
     \sim \Bi \left (n_{tkz},  \logitInv  \left \{ \beta_{tz, [2]} + u_2  \right \}  \right ) 
      ,\, \forall u_2\in\re 
     \\
     Y_{tkz}^3 \vert U_{tk}^3 = u_3 
     \sim \Bi \left (n_{tkz},  \logitInv  \left \{ \beta_{tz, [3]}+ u_3  \right \} \right )
      ,\, \forall u_3\in\re 
     \\
     C_{tkz}^1 \vert V_{tk}^1=v_1
      \sim
      \Po \left ( \exp \left \{  \theta_{tz, [1]} + v_1     \right \}  \right )
       ,\, \forall v_1\in\re  
      \\   
     C_{tkz}^2 \vert V_{tk}^2=v_2
      \sim
      \Po \left ( \exp \left \{  \theta_{tz, [2]} + v_2     \right \}  \right )
      ,\, \forall v_2\in\re 
     \\
     C_{tkz}^3 \vert V_{tk}^3=v_3
      \sim
      \Po \left ( \exp \left \{  \theta_{tz, [3]} + v_3     \right \}  \right ) \, 
       ,\, \forall v_3\in\re \, .
 \end{array}
 \right.
\end{align}
% for all $u_1, u_2, u_3, v_1, v_2, v_3$ in $\re$.
Here $U_{tk}\dd$ and $V_{tk}\dd$ (for $d=1,2,3$, $t=1,\ldots,4$ and $k=1,\ldots,6$) are Gaussian random component representing the  $k\tth$ tube exposed to the $t\tth$ treatment belonging to a marginal model describing the  \emph{scatter} and the \emph{intensity}, respectively, at the $d\tth$ development stage.
We assume that 
$(U_{tk}^{[1]},U_{tk}^{[2]},U_{tk}^{[3]},V_{tk}^{[1]},V_{tk}^{[2]}, V_{tk}^{[3]})$ is multivariate normally distributed with expectation zero and covariance matrix $\bm{\Sigma}$. Furthermore, we assume that  
$(U_{tk}^{[1]},U_{tk}^{[2]},U_{tk}^{[3]},V_{tk}^{[1]},V_{tk}^{[2]},$ $V_{tk}^{[3]} )$  and $(U_{t'k'}^{[1]},U_{t'k'}^{[2]},U_{t'k'}^{[3]},$ $V_{t'k'}^{[1]},V_{t'k'}^{[2]},V_{t'k'}^{[3]})$ are independent when  $(t,k) \ne (t',k')$, \ie, we assume that the components of the random components corresponding to different tubes are independent.

\subsection{Modelling the Covariance Structure of the Random Components}\label{SubSection.3.2}

The next step in the construction of the multivariate model we have in mind is to model the covariance structure of the random components by a graphical model. 
Before embracing this project, we give a short account of the theory of graphical models.
For a comprehensive description see \cite{Lauritzen1996} and \cite{Whittaker1990}.

Consider a graph $\mathcal{G}=(\mathcal{V}, \mathcal{E})$ with vertices composed of random variables. The set of edges $\mathcal{E} \subseteq \mathcal{V}\times \mathcal{V}$ is formed with the convention that two vertices are connected by an edge if, and only if, they are not conditionally independent given the remaining random variables in $\mathcal{V}$. We say that there is a path between two vertices, say $V_1$ and $V_2$, if there exist a sequence of pairs of vertices in $\mathcal{E}$
such that  $V_1$ and $V_2$ belong to at least one vertex of the sequence. 
A set of vertices $\mathcal{S}$ is said to separate the sets of vertices $\mathcal{A}$ and $\mathcal{B}$ in the graph, if and only if, each path connecting an element of $\mathcal{A}$ to an element of $\mathcal{B}$ contains at least one element of $\mathcal{S}$.
The separation principle (or global Markov property) is a crucial property of graphical models stating that if a set of vertices, $\mathcal{S}$, separates two disjoint subsets of vertices $\mathcal{A}$ and $\mathcal{B}$, then all variables in $\mathcal{A}$ are independent of all variables in $\mathcal{B}$ given the variables in $\mathcal{S}$, see \cite{Lauritzen1996} and \cite{Perl2009}.

In the construction of the multivariate model in question here, we consider a graph
 $\mathcal{G}=(\mathcal{V}, \mathcal{E})$  with vertices
$\mathcal{V} = \left\{ U^{[1]},U^{[2]},U^{[3]},V^{[1]},V^{[2]},V^{[3]}\right \}$
formed by the random variables representing the random components of the multivariate GLMM described in section \ref{SubSection.3.1} (with the obvious notational convention, \eg,  $U^{[1]}$ is the random variable with the same distribution  as $U_{tk}^{[1]}$, for $t=1,\ldots, 4$, $k=1,\ldots,6$). 
Clearly, $ \left ( U^{[1]},U^{[2]},U^{[3]},V^{[1]},V^{[2]},V^{[3]}\right ) \sim 
   N\left (\bm{0}, \bm{\Sigma} \right )$ by construction. 
Here we define the set of edges $\mathcal{E}$ using the conditional independence of pairs of random variables as in the paragraph above. 

The interpretation of the graph above in terms of the random components of the multivariate generalised mixed model defined in section \ref{SubSection.3.1} is straigtforward. For example, if there is an edge connecting $U^{[1]}$ and
$V^{[1]}$, then these two random variables are not conditionally independent given the other random components; therefore, $U^{[1]}$ carries some information on $V^{[1]}$ and this information is not contained in the other random components. 
Note that  $U^{[1]}$ and $V^{[1]}$ represent a latent variation on the \emph{scatter} and the \emph{intensity}, respectively, after having corrected for the effects of the treatment and the depth zone. 
Therefore, we would conclude that we have evidence that some latent mechanisms governing   the \emph{scatter} and \emph{intensity} are related. The strength of this association can be estimated by inferring the entry of the precision matrix (\ie, $\bm{\Sigma}^{-1}$) corresponding to the random components $U^{[1]}$ and $V^{[1]}$. 

Note that the separation principle also holds in this graph, which is crucial for the interpretation of the model in terms of the covariance structure of the random components. Moreover, it is possible to extend the separation principle, obtaining what we call the \emph{induced separation principle},  to draw some general conclusions on the response variables, as we explain below using a putative example. 
Consider the groups of random components $\mathcal{A} = \left \{ U^{[1]}, V^{[1]} \right \}$ and $\mathcal{B} = \left \{ U^{[3]}, V^{[3]} \right \}$ and $\mathcal{S} = \left \{ U^{[2]}, V^{[2]} \right \}$ contained in  $\mathcal{V}$. 
Define (for any choice of $t=1,\ldots, 4$, $k=1,\ldots,6$, $z=A, B ,C$)
the sets of response variables 
$\widetilde{\mathcal{A}}  = \left \{ Y_{tkz}^{[1]}, C_{tkz}^{[1]} \right \}$ and 
$\widetilde{\mathcal{B}} =  \left \{ Y_{tkz}^{[3]}, C_{tkz}^{[3]} \right \} $. 
The sets $\widetilde{\mathcal{A}}$ and $\widetilde{\mathcal{B}}$ are the sets of response variables of the marginal models for which the elements of $\mathcal{A}$ and $\mathcal{B}$ are random components. 
According to the induced separation principle, if  $\mathcal{S}$ separates $\mathcal{A}$ and $\mathcal{B}$ in the graph $\mathcal{G}=(\mathcal{V}, \mathcal{E})$, then the random variables in $\widetilde{\mathcal{A}}$ are conditionally independent of the random variables in $\widetilde{\mathcal{B}}$ given the random variables in $\mathcal{S}$. The proof of the induced separation principle can be done by using basic properties of conditional densities and using the factorisation of the joint densities of the distributions of $\mathcal{V}$, see the details in \cite{Pelck2020B}. 

We stress that in the putative example above, conditioning on the responses in $\widetilde{\mathcal{S}} =  \left \{ Y_{tkz}^{[2]}, C_{tkz}^{[2]} \right \}$ (\ie, the corresponding response variables to the elements of $\mathcal{S}$) does not necessarily render the random variables in $\widetilde{\mathcal{A}}$ independent of the random variables in $\widetilde{\mathcal{B}}$.
Still in the putative example in discussion, the fact that the group of random components  $\mathcal{S}$ separates  $\mathcal{A}$ and  $\mathcal{B}$ implies, in the present setup, that the responses at the first development stage are conditionally independent of the responses at the third development stage, given the random components related to the second development stage. We will see in section \ref{Section.4} that this indeed the case. 

\subsection{Inferring the Covariance Structure}\label{SubSection.3.3}

The graph  $\mathcal{G}=(\mathcal{V}, \mathcal{E})$ with vertices
$\mathcal{V} = \left\{ U^{[1]},U^{[2]},U^{[3]},V^{[1]},V^{[2]},V^{[3]}\right \}$ defining the covariance structure of the random components introduced in section \ref{SubSection.3.2} can be inferred by predicting the  random components and using those predictors to infer a graphical model that minimises the BIC, as proposed and implemented in \cite{Abreu2010} and \cite{Edwards2010}. 
The predictors of the random components of the generalised linear mixed models can be obtained with inference procedures yielding consistent and normally distributed predictors. We used the procedure described in \cite{Pelck2020A}. Additionally, we tested each of the possible vertices using the conditional test for random components under multivariate generalised linear mixed models described in \cite{Pelck2020B}.

\section{Analysing the Motivational Example}\label{Section.4}

\begin{figure}[!htbp]
	\centering
	\includegraphics[width=1.0\textwidth]{"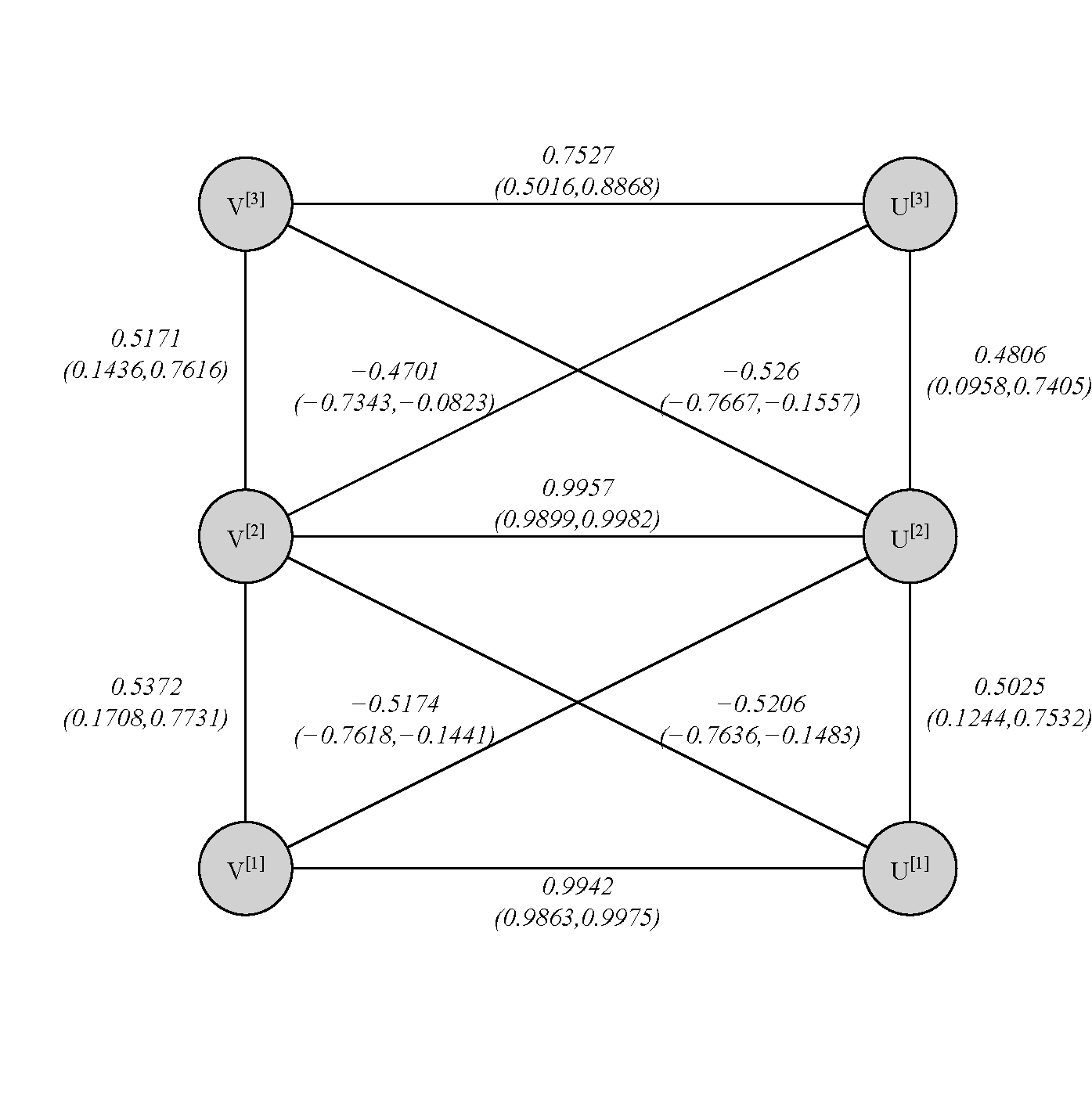"}
	\caption{Representation of the graph  
	               $\mathcal{G}=(\mathcal{V}, \mathcal{E})$  with vertices
                        $\mathcal{V} = \left\{ U^{[1]},U^{[2]},U^{[3]},V^{[1]},V^{[2]},V^{[3]}\right \}$
         formed by the random variables representing random components of the multivariate GLMM 
         discussed in section \ref{Section.4}. 
         The graph was estimated by minimising the BIC of a covariance selection model inferred using the 
         prediction of the random components.
	The super-indices in the vertices indicate that which development stage the response is observed and the letter $V$ 
	indicates that the random component correspond to the \emph{intensity} while $U$ represents the random components associated to the \emph{scatter}. 
	The numbers placed besides the edges are the estimated partial correlations with the corresponding confidence intervals (with coverage of 0.95).  
	}\label{fig:graph}
\end{figure}

The example described in Section \ref{SubSection.2.1} is analysed below. Figure \ref{fig:graph} displays a representation of the  graph $\mathcal{G}=(\mathcal{V}, \mathcal{E})$ with vertices
$\mathcal{V} = \left \{ U^{[1]},U^{[2]},U^{[3]},V^{[1]},V^{[2]},V^{[3]}\right \}$ estimated by minimising the BIC. Additionally, we tested whether the conditional covariances of each of the possible pairs of elements of $\mathcal{V}$, given the other elements of $\mathcal{V}$, is zero. The conditional covariances corresponding to the edges in the graph $\mathcal{G}$ displayed in  Figure \ref{fig:graph} were all significantly different than zero (at a significance level of $0.05$). Moreover, the conditional covariances corresponding to the pairs of elements of  $\mathcal{V}$ that are not in  $\mathcal{E}$ were not significantly different than zero (at a significance level of $0.05$). 

According to the graph $\mathcal{G}=(\mathcal{V}, \mathcal{E})$ displayed in Figure \ref{fig:graph}, the \emph{scatter} and the \emph{intensity} are positively conditionally correlated at each of the three development stages studied, suggesting the presence of underlying mechanisms associated to the \emph{scatter} and the \emph{intensity} that are positively associated. The information that \emph{scatter} carries on the \emph{intensity} (and vice-versa), for each development stage, are not contained in the other random components, \ie, there is evidence of specific common or cooperative underlying mechanisms determining the \emph{scatter} and the \emph{intensity}, specific for each development stage.
This result rules out the possibility that the plants would be compensating a reduced occupation of the soil by increasing the intensity of the colonisation of the soil by the radicular system. Moreover, the strength of these associations is essentially the same in the first two development stages (testing the equality of the two conditional correlations yields the p-value $0.979$), but decreases in the last development stage (p-values for comparing the first and the third stage and the second and the third stage are $<0.001$ and $<0.0001$, respectively).

The inferred graph $\mathcal{G}$ indicates  the presence of a temporal Markovianity in the sense that the random components related to the first and the third development stages are conditionally independent given the random components associated to the second development stage. To see that, note that the set $\mathcal{S} = \left \{ U^{[2]},V^{[2]} \right \}$ separates the sets  $\mathcal{A} = \left \{ U^{[1]},V^{[1]}\right \}$ and $\mathcal{B} = \left \{ U^{[3]},V^{[3]} \right \}$ in the graph $\mathcal{G}$. It follows then from the separation principle that the random components of $\mathcal{A}$ are conditionally independent of the random components of $\mathcal{B}$ given $\mathcal{S}$. 

Using the induced separation principle, we conclude that for any choice of treatment $t=1,\ldots, 4$, tube $k=1,\ldots,6$, and depth zone $z=A, B ,C$, the group of responses $\widetilde{\mathcal{A}} = \left \{ Y_{tkA}^{[1]},C_{tkA}^{[1]}, Y_{tkB}^{[1]},C_{tkB}^{[1]}, Y_{tkC}^{[1]},C_{tkC}^{[1]}\right \}$ and $\widetilde{\mathcal{B}} = \left \{ Y_{tkA}^{[3]},C_{tkA}^{[3]}, Y_{tkB}^{[3]}\right.$ $\left. ,C_{tkB}^{[3]}, Y_{tkC}^{[3]},C_{tkC}^{[3]}\right \}$ are conditional independent given the group of random components $\mathcal{S} = \left \{ U^{[2]},V^{[2]} \right \}$. That is, all the information that the response variables observed in the first development stage  (\ie, the variables in  $\widetilde{\mathcal{A}}$)  might carry on the response variables in the third stage  (\ie, the variables in  $\widetilde{\mathcal{B}}$) is entirely contained in the informational contents contained in the random components associated to the second development stage, and this conclusion holds for any combination of  fertilisation treatment and depth zone. This lack of memory result described above indicates that there are no long lasting temporal causal effects affecting the roots' development (in terms of  the \emph{scatter} and the \emph{intensity}), which is a non-trivial conclusion that cannot be reached with univariate models for describing the \emph{scatter} and the \emph{intensity}.

\clearpage

\section{Discussion and Conclusion}\label{Section.5}

In this article, we combine multivariate GLMMs and graphical models (GMs) to characterise the spatial and temporal development of roots systems in a cultivation field. The use of graphical models in biological and agricultural research is not new, see \cite{Labouriau2000}, \citet[see the SOM]{Labouriau2008}, \cite{Holmstrup2011}, \cite{Baral2017} and \cite{Taghizadeh-Toosi2019} were GMs are applied in different agricultural and biological contexts. In all these applications, the nature of the multivariate responses is essentially the same in the sense that whether the responses are continuous multivariate normally distributed or discrete multivariate multinomial distributed. 

In \cite{Lamande2011} and \citet[using different data from the same experiment described here]{Azeez2020}, the GM is of mixed type involving continuous multivariate normal distributed responses and discrete multivariate multinomial distributed responses. There, the GM uses the so-called CG (Conditional Gaussian) distributions as defined in \cite{Edwards2010} see also \cite{Abreu2010} for implementation and further discussions. However, in the current application, it is not possible to use multinomial based- graphical models since the supports of the distribution of counts (used to model the \emph{intensity}) are not bounded and the binomial distributions (used for modelling the \emph{scatter}) have different sizes for different observations. Therefore, one cannot do the analysis we discussed here using standard graphical models. The price we pay for modelling responses of varying nature (using graphical models  that are not based on the CG distribution) is that we are forced to use random components, whether by directly interpreting the random components or by using the induced separation principle. 

From the applied point of view, this article exemplifies the use of a combination of GLMMs and GMs, which allows exploring biological aspects such as the temporal development of two quantities of different nature. Here the close interface between Statistics and Biology played a crucial role in the modelling process. 

% ============================================================= %
\section*{Acknowledgements}
% ============================================================= %
\Fnine
We thank Hanne Lakkenborg Kristensen for having introduced us to the use of minirhizotron for characterising the development of the radicular system in fields. The statistical modelling presented here is a fruit of discussions with Hanne Lakkenborg Kristensen, Gitte Holton Rub{\ae}k, Lars Juhl Munkholm, Margita Hefner, Julie Therese Christensen, and Ellen Margrethe Wahlstr\" {o}m, all of them from the Aarhus University. Gitte Holton Rub{\ae}k kindly supplied us the data of the motivational example.
 The first author was partially financed by the Applied Statistics Laboratory (aStatLab) at the Department of Mathematics, Aarhus University. 
 \normalsize
\newpage

% ------------------------------------------------------------------------

\newpage

\appendix
\section{Representation of the Covariance Structure in Terms of Direct Acyclic Graphs}
The covariance structure of the responses and the random components at different developing stages can be represented in terms of a Direct Acyclic Graph (DAG) as follows. Here we use the terminology and properties of graphical models represented as DAGs exposed in  \cite{Lauritzen1996}. This representation follows from the induced separation principle exposed in Section \ref{SubSection.3.2}  and the properties of the GLMM. 

\begin{figure}[htbp]
    \begin{center} \scalebox{1.0}{
                                  \begin{tikzpicture}[
                                                                                                                    squarednode/.style={rectangle, draw=black, fill=lightgray, thick},
                                                                                                                    roundnode/.style={circle, draw=black, fill=lightgray, thick},
                                                                                                                    ]
                                                                                                                    \node [style=roundnode] (V3) at (4, 12) {$V_{tk}^{[3]}$};
                                                                                                                    \node [style=roundnode] (V1) at (4, 4) {$V_{tk}^{[1]}$};
                                                                                                                    \node [style=roundnode] (U1) at (8, 4) {$U_{tk}^{[1]}$};
                                                                                                                    \node [style=roundnode] (V2) at (4, 8) {$V_{tk}^{[3]}$};
                                                                                                                    \node [style=roundnode] (U2) at (8, 8) {$U_{tk}^{[2]}$};
                                                                                                                    \node [style=roundnode] (U3) at (8, 12) {$U_{tk}^{[3]}$};
                                                                                                                    \node [style=squarednode] (C11) at (2, 5) {$C_{tkA}^{[1]}$};
                                                                                                                    \node [style=squarednode] (C12) at (2, 4) {$C_{tkB}^{[1]}$};
                                                                                                                    \node [style=squarednode] (C13) at (2, 3) {$C_{tkC}^{[1]}$};
                                                                                                                    \node [style=squarednode] (C21) at (2, 9) {$C_{tkA}^{[2]}$};
                                                                                                                    \node [style=squarednode] (C22) at (2, 8) {$C_{tkB}^{[2]}$};
                                                                                                                    \node [style=squarednode] (C23) at (2, 7) {$C_{tkC}^{[2]}$};
                                                                                                                    \node [style=squarednode] (C31) at (2, 13) {$C_{tkA}^{[3]}$};
                                                                                                                    \node [style=squarednode] (C32) at (2, 12) {$C_{tkB}^{[3]}$};
                                                                                                                    \node [style=squarednode] (C33) at (2, 11) {$C_{tkC}^{[3]}$};
                                                                                                                    \node [style=squarednode] (Y11) at (10, 5) {$Y_{tkA}^{[1]}$};
                                                                                                                    \node [style=squarednode] (Y12) at (10, 4) {$Y_{tkB}^{[1]}$};
                                                                                                                    \node [style=squarednode] (Y13) at (10, 3) {$Y_{tkC}^{[1]}$};
                                                                                                                    \node [style=squarednode] (Y21) at (10, 9) {$Y_{tkA}^{[2]}$};
                                                                                                                    \node [style=squarednode] (Y22) at (10, 8) {$Y_{tkB}^{[2]}$};
                                                                                                                    \node [style=squarednode] (Y23) at (10, 7) {$Y_{tkC}^{[2]}$};
                                                                                                                    \node [style=squarednode] (Y31) at (10, 13) {$Y_{tkA}^{[3]}$};
                                                                                                                    \node [style=squarednode] (Y32) at (10, 12) {$Y_{tkB}^{[3]}$};
                                                                                                                    \node [style=squarednode] (Y33) at (10, 11) {$Y_{tkC}^{[3]}$};
                                                                                                                    \draw [-] (V1) to (U1);
                                                                                                                    \draw [-] (V2) to (U2);
                                                                                                                    \draw [-] (V3) to (U3);
                                                                                                                    \draw [-] (V1) to (V2);
                                                                                                                    \draw [-] (V2) to (V3);
                                                                                                                    \draw [-] (U1) to (U2);
                                                                                                                    \draw [-] (U2) to (U3);
                                                                                                                    \draw [-] (V1) to (U2);
                                                                                                                    \draw [-] (U1) to (V2);
                                                                                                                    \draw [-] (V2) to (U3);
                                                                                                                    \draw [-] (V3) to (U2);
                                                                                                                    \draw [->] (V1) to (C11);
                                                                                                                    \draw [->] (V1) to (C12);
                                                                                                                    \draw [->] (V1) to (C13);
                                                                                                                    \draw [->] (V2) to (C21);
                                                                                                                    \draw [->] (V2) to (C22);
                                                                                                                    \draw [->] (V2) to (C23);
                                                                                                                    \draw [->] (V3) to (C31);
                                                                                                                    \draw [->] (V3) to (C32);
                                                                                                                    \draw [->] (V3) to (C33);
                                                                                                                    \draw [->] (U1) to (Y11);
                                                                                                                    \draw [->] (U1) to (Y12);
                                                                                                                    \draw [->] (U1) to (Y13);
                                                                                                                    \draw [->] (U2) to (Y21);
                                                                                                                    \draw [->] (U2) to (Y22);
                                                                                                                    \draw [->] (U2) to (Y23);
                                                                                                                    \draw [->] (U3) to (Y31);
                                                                                                                    \draw [->] (U3) to (Y32);
                                                                                                                    \draw [->] (U3) to (Y33);
                                                                                       \end{tikzpicture}
                                                          }
                             \end{center}
                             \caption{DAG representation of the covariance structure of the responses and the random components for the $tk\tth$ tube at different developing stages induced by the graph  $\mathcal{G}=(\mathcal{V}, \mathcal{E})$.  Edges without arrows represent arrows in both directions. The structure disputed holds for any choice of  treatment $t$ ($t=1,\ldots, 4$) and tube $k$ ($k=1,\ldots,6$). }
 \label{fig:graph2}                          
\end{figure}
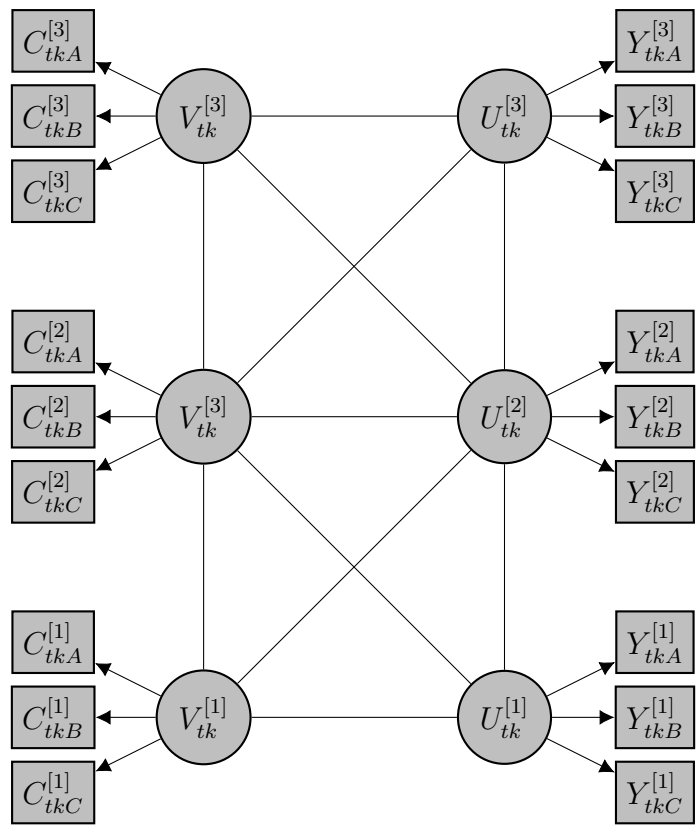

According to the inferred multivariate model,  for any possible choice of treatment $t=1,\ldots, 4$, tube $k=1,\ldots,6$, and depth zone $z=A, B ,C$,  the covariance structure of the responses and random components is expressed by the DAG represented in Figure \ref{fig:graph2}. 

\end{document}